   \documentclass{aa}
\usepackage{graphicx}
\usepackage{rotating}
\def\gtrsim{\mathrel{\hbox{\rlap{\hbox{\lower4pt\hbox{$\sim$}}}\hbox{$>$}}}}
\def\ltsim{\mathrel{\hbox{\rlap{\hbox{\lower4pt\hbox{$\sim$}}}\hbox{$<$}}}}
 \begin{document}
\title{Variations of the high-level Balmer line spectrum of the helium-strong star $\sigma$\,Orionis\,E\thanks{Based on observations obtained at the
        the Dominion Astrophysical Observatory, Herzberg Institute of Astrophysics,
        National Research Council of Canada.}}
\authorrunning {Smith \& Bohlender }
\titlerunning{High-level Balmer lines in $\sigma$\,Ori\,E}
\author{M.A. Smith\inst{1} and D.A. Bohlender\inst{2} }
\institute{Department of Physics, Catholic University of America,
Washington, DC 20064, USA; Present address: Space Telescope Science
Institute, 3700 San Martin Dr., Baltimore, MD 21218 ~~Email:~ msmith@stsci.edu \and
National Research Council of Canada, Herzberg Institute of Astrophysics, 5071 W. Saanich Rd., Victoria, BC Canada V9E 2E7 }

\date{Received ??; accepted ??}
\abstract{
%Context:
{}
{Using the high-level Balmer lines and continuum,
we trace the density structure of two magnetospheric
disk segments of the prototypical Bp star $\sigma$\,Orionis\,E (B2p) as
these segments occult portions of the star during the rotational cycle.  }
{High-resolution spectra of the Balmer lines $\ge$H9 and Balmer edge were
obtained on seven nights in January-February 2007 at an average sampling
of 0.01 cycles. We measured equivalent width variations due to the star
occultations by two disk segments 0.4 cycles apart and constructed
differential spectra of the migrations of the corresponding absorptions
across the Balmer line profiles. We first estimated the rotational and
magnetic obliquity angles. We then simulated the observed Balmer jump
variation using the model atmosphere codes \mbox{\it synspec/circus} and
evaluated the disk geometry and gas thermodynamics.
}
{We find that the two occultations are caused by two disk segments.  The
first of these transits quickly, indicating that the 
segment resides in a range of distances, perhaps 2.5$-$6R$_{*}$, from 
the star.  The second consists of a more slowly moving segment situated 
closer to the surface and causing two semi-resolved absorbing maxima.   
During its transit this segment brushes across the star's ``lower" limb. 
Judging from the line visibility up to H23-H24 during the
occultations, both disk segments have mean densities near 10$^{12}$
cm$^{-3}$ and are opaque in the lines and continuum. They have semiheights 
less than ${\frac 12}$R$_{*}$, and their temperatures are near 10\,500\,K 
and 12\,000\,K,
respectively. 
%Outside the magnetic plane lies static gas with a density of approximately
}
{In all, the disks of Bp stars have a much more complicated geometry
than has been anticipated, as evidenced by their (sometimes)
non-coplanarity, de-centerness, and from star to star, differences in
disk height.
}
\keywords{stars: individual: $\sigma$\,Orionis\,E --
stars: magnetic fields -- stars: circumstellar matter -- stars: winds, outflows
-- stars: early-type -- ultraviolet: stars } }
\maketitle

\section{Introduction}

   Over the last several years the interest in the class of magnetic Bp 
stars has spurred the accumulation of datasets at many wavelengths over
the stars' rotational/magnetic cycles. These datasets support the view 
that the circumstellar environment is the site of a balance of physical 
processes that channels the wind toward the magnetic plane. If the magnetic
field is strong, that is if the circumstellar magnetic energy density 
exceeds the wind energy density, the star's wind is channeled into 
the magnetic plane where a shock is formed and the cooled shock settles
to form a stable disk. The magnetic axis is typically inclined with respect
to the rotational axis. Then, as different portions of the disk wobble in 
front of the star during the rotational cycle, they produce alternate 
absorption and emission components in the H$\alpha$, UV resonance lines,
and low-excitation metallic lines  (e.g., Bohlender et al. 1987, Shore
1987, Bolton 1994). 
To complicate the picture, it appears that slow infall of disk matter
is responsible for producing inhomogeneous metal-poor patches or a 
belt along the magnetic equator (e.g., Khokhlova et al. 2000, 
Groote 2003). Also, although there remains a theoretical difficulty with the
fractionation hypothesis for helium (Krti\v{c}ka et al. 2006), 
enough neutral helium atoms in fact seem likely to separate from the polar 
wind, return to the surface, and accumulate there as helium caps. 
  
  Recent theoretical refinements in the picture (e.g., Preuss et 
al. 2004, Townsend \& Owocki 2005; hereafter TO05) suggest that the wind 
will settle onto low equipotential surfaces determined by the balance 
of radiative, gravitational, magnetic, and centrifugal forces. 
These surfaces 
reside primarily, though not exclusively, in the disk plane. Because
the disk is likely to have a nonaxisymmetric density distribution, we
will refer to the accumulations near the plane responsible for 
time-variable spectral features as disk segments. TO05 have presented 
a ``rigidly rotating magnetospheric" {\it ab initio} model to explain 
physical properties of these segments for the case of $\sigma$\,Ori\,E
and it can be applied to the cases of other Bp stars with arbitrary
magnetic inclinations and viewed from any rotational inclinations.
TO05 find that the disk plane is not precisely
perpendicular to the magnetic polar axis for intermediate magnetic
obliquities, including those they obtained for $\sigma$\,Ori\,E.

  The spectroscopic anomalies of  $\sigma$\,Ori\,E (HD\,37479, B2\,Vp; 
Lesh 1968) were first noticed by Berger (1956) and discussed further by
Osmer and Peterson (1972). This star 
is the prototype and most studied of the helium-strong subclass of Bp stars.
Walborn (1974) discovered H$\alpha$ emissions in this star's spectrum 
that varied on a timescale of hours. This variability was subsequently 
established to be associated with the rotation of a frozen-in disk, as 
described above. Hesser et al. (1977) determined a period of 1.19081 days
based on several nights of photometric observations distributed over a few
years. Subsequently, various studies
(Landstreet \& Borra 1978, Bohlender et al. 1987) determined that the
magnetic variations (suggestive of an approximately
dipolar field with polar field strength B$_p \approx$ 10\,kG) 
are consistent with the photometric
period. Various periods over the range
1.19081$-$1.19084 days have since been determined by several authors from
analyses of variations of photospheric and circumstellar H$\alpha$ line
emissions
(Hesser, Moreno, \& Ugarte 1977, Reiners et al. 2000, hereafter R00; 
Townsend, Owocki, \& Groote  2005; hereafter TOG05). 

Groote \& Hunger (1976, 1977 hereafter GH76 and GH77) discussed the variations 
of high-level Balmer lines based on only photographic spectra. 
The GH76 data indicated that two disk occultations  of the star, 
about 0.4 cycles apart, cause high-level Balmer line absorptions visible 
at these times. At the same times, the intervening 
disk material absorbs flux just shortward of the Balmer edge. Although the 
analysis of line absorptions do not generally allow us to infer the extent 
and total volume of such a corotating structure, these absorptions do yield 
considerable information about its three dimensional geometrical character. 
Smith \& Groote (2001, hereafter SG01) combined 
strength and shape information of many ultraviolet metallic lines obtained 
by the {\it International Ultraviolet Explorer} around the rotation cycle to
exploit this potential. Their study provides some of the first estimates
of the physical parameters of the disk gas, including disk temperature,
column density, and areal coverage of the star.  The limitations of this
study were the low signal-to-noise ratio and the paucity of phase sampling
of the archival database. What has been needed is a study that 
includes complete rotational phase coverage  of absorption lines sensitive 
to electron density.  To respond to this need, we have obtained complete phase
coverage of the hydrogen lines from H9 (H$\zeta$) to the Balmer edge over 
the entire rotation period of $\sigma$\,Ori\,E. Such datasets have the
potential to better elucidate 
areal coverage and density structure of the absorbing
disk segments than has previously been possible. 

\section{Observations and modeling methodology }

\subsection{Observations}

In an attempt to search for variations in high-level Balmer lines
from the variable occultation of $\sigma$\,Ori\,E by its circumstellar disk, we obtained 114
spectra with the 1.85\,m Plaskett telescope of the Dominion Astrophysical Observatory (DAO)
on seven nights between 2007 January 25 and February 28. 
We used the f/5 Cassegrain spectrograph with an 1800 l/mm grating blazed at 
5000\AA\/ and the SITe-2 CCD.
Despite the sometimes mediocre seeing at the DAO, we used a narrow 1\,arcsec slit rather than an available image slicer in order to maximize
the resolution and to permit background sky subtraction.  
We note, however, that observing conditions and seeing were very good on each of our observing nights.
Exposure times were 15 minutes and to minimize the effects of any possible 
flexure in the spectrograph Fe/Ar comparison arc reference spectra were 
obtained approximately every hour.  The data were processed 
in a standard fashion with IRAF.\footnote{IRAF is distributed by the 
National Optical Astronomy Observatory, which is operated by the 
Association of Universities for Research in Astronomy (AURA), Inc., under 
cooperative agreement with the National Science Foundation.}
The resulting spectra have a
dispersion of 10 \AA\,mm$^{-1}$, a resolution of about 16\,400, and
cover the wavelength region $\lambda\lambda$3645$-$3863. 
Typical signal-to-noise ratios,
as measured in a flat region of the spectrum near $\lambda$3850, are
about 150 per pixel.  Table\,1 provides a journal
of observation numbers (the DAO 1.8-m odometer numbers for 2007 spectra), observing times expressed as Heliocentric Julian Day (HJD) and rotational phases.

\begin{table*}[ht!]
\centering
\caption{$\sigma$\,Ori\,E observing log including DAO 1.8-m spectrograph 2007 
odometer numbers. Heliocentric Julian Dates - 2450000.0 and rotation phases are
given in columns 2 and 3. The phases are derived using a  formal period of
1.190841 days and the 1976 zeropoint of HJD\,2442778.819 used by Hesser et al. 
(1977)  - see text.} 
\begin{tabular}{c c c| c c c | c c c | c c c }  \hline
 \hline \\ [-2.2ex]
Obsn. & HJD & $\phi$   & Obsn. & HJD & $\phi$ & Obsn. & HJD  & $\phi$ & Obsn. & HJD  & $\phi$ \\
\# & & &\# & & &\# & &  &\# & & \\
 \hline
 069 &   4125.724 &       0.480 &        373 &   4130.696 &       0.655 &        453 &   4131.774 &       0.561 &        637 &   4133.674 &       0.156  \\
 072 &   4125.748 &       0.500 &       375 &   4130.709 &       0.666   &454 &   4131.785 &       0.569 &        638 &   4133.684 &       0.165  \\
 073 &   4125.758 &       0.509 &       376 &   4130.720 &       0.675    &456 &   4131.796 &       0.579 &        639 &   4133.695
 &       0.174  \\ 
 074 &   4125.769 &       0.517 &       377 &   4130.730 &       0.684    &457 &   4131.807 &       0.588 &        641 &   4133.707
 &       0.184  \\
 076 &   4125.783 &       0.529 &       378 &   4130.741 &       0.693    &508 &   4132.622 &       0.272 &        642 &   4133.718
 &       0.192  \\ 
 077 &   4125.794 &       0.538 &       380 &   4130.752 &       0.702    &509 &   4132.632 &       0.281 &        643 &   4133.728
 &       0.201  \\
 078 &   4125.804 &       0.547 &       381 &   4130.763 &       0.711    &510 &   4132.643 &       0.290 &        644 &   4133.738
 &       0.210  \\ 
 079 &   4125.814 &       0.556 &       382 &   4130.773 &       0.720    &511 &   4132.653 &       0.298 &        646 &   4133.750
 &       0.220  \\
 298 &   4129.672 &       0.795 &       383 &   4130.784 &       0.729    &513 &   4132.665 &       0.308 &        647 &   4133.761
 &       0.229  \\
 299 &   4129.682 &       0.804 &       385 &   4130.796 &       0.739    &514 &   4132.675 &       0.317 &        648 &   4133.771
 &       0.238  \\ 
 300 &   4129.693 &       0.813 &       386 &   4130.807 &       0.748    &515 &   4132.686 &       0.326 &        649 &   4133.782
 &       0.246  \\
 301 &   4129.703 &       0.822 &       387 &   4130.817 &       0.757    &516 &   4132.696 &       0.335 &        651 &   4133.794
 &       0.256  \\
 303 &   4129.715 &       0.831 &       388 &   4130.828 &       0.766    &518 &   4132.708 &       0.345 &        652 &   4133.804
 &       0.265  \\
 304 &   4129.726 &       0.840 &       389 &   4130.840 &       0.776    &519 &   4132.719 &       0.354 &        653 &   4133.814
 &       0.274  \\
 305 &   4129.736 &       0.849 &       390 &   4130.850 &       0.784    &520 &   4132.729 &       0.362 &        654 &   4133.825
 &       0.283  \\
 306 &   4129.747 &       0.858 &       436 &   4131.623 &       0.434    &521 &   4132.740 &       0.371 &        795 &   4159.640
 &       0.960  \\
 308 &   4129.758 &       0.868 &       437 &   4131.634 &       0.443    &523 &   4132.751 &       0.381 &        796 &   4159.651
 &       0.970  \\
 309 &   4129.769 &       0.877 &       438 &   4131.644 &       0.451    &524 &   4132.762 &       0.390 &        797 &   4159.662
 &       0.979  \\
 310 &   4129.779 &       0.885 &       439 &   4131.655 &       0.460    &525 &   4132.772 &       0.399 &        798 &   4159.672
 &       0.988  \\
 311 &   4129.790 &       0.894 &       441 &   4131.667 &       0.470    &526 &   4132.783 &       0.407 &        800 &   4159.687
 &       0.000  \\
 313 &   4129.802 &       0.905 &       442 &   4131.677 &       0.479    &528 &   4132.794 &       0.417 &        801 &   4159.697
 &       0.009  \\
 314 &   4129.813 &       0.914 &       443 &   4131.688 &       0.488    &529 &   4132.805 &       0.426 &        802 &   4159.708
 &       0.018  \\
 315 &   4129.823 &       0.922 &       444 &   4131.698 &       0.497    &530 &   4132.815 &       0.435 &        803 &   4159.718
 &       0.026  \\
 316 &   4129.834 &       0.931 &       446 &   4131.710 &       0.507    &531 &   4132.826 &       0.443 &        805 &   4159.731
 &       0.037  \\
 317 &   4129.844 &       0.940 &       447 &   4131.720 &       0.515    &631 &   4133.620 &       0.111 &        806 &   4159.741
 &       0.045  \\
 319 &   4129.856 &       0.950 &       448 &   4131.731 &       0.524    &632 &   4133.631 &       0.120 &        807 &   4159.751
 &       0.054  \\
 370 &   4130.665 &       0.629 &       449 &   4131.741 &       0.533    &633 &   4133.641 &       0.128 &        808 &   4159.762
 &       0.063  \\
 371 &   4130.675 &       0.638 &       451 &   4131.753 &       0.543    &634 &   4133.652 &       0.137 &            &          
 &              \\
 372 &   4130.686 &       0.647 &       452 &   4131.764 &       0.552    &636 &   4133.664 &       0.147 &            &          
 &              \\
\hline
\end{tabular}
%\end{center}
\end{table*}

\subsection{Spectral synthesis computations}

  To perform the analysis of the high-level Balmer line and continuum
spectra, we utilized a pair of programs written by I. Hubeny and T. Lanz
with both LTE (Kurucz 1993) and non-LTE models in the BSTAR2006 series
(Lanz \& Hubeny 2007). The first of these, {\it synspec,} is a photospheric
line synthesis program that computes
the line and continuum photospheric spectrum of a star
(Hubeny, Lanz, \& Jeffery 1994).
We also made use of {\it synspec}'s capability of convolving the photospheric 
lines with functions approximating the instrumental and rotational broadening.  We computed the synthetic spectra in steps of 0.01\AA.

 The goal of our analysis was to compute effects of the high-level
Balmer lines and Balmer continuum due to the intervening disk during 
each of the two occultation events that occur during $\sigma$\,Ori\,E's rotation cycle. 
Since it is already
known from the SG01 study that the disk is cooler than the photosphere,
these effects will only include absorptions. To simulate these absorptions we used 
the radiative transfer program {\it circus} (Hubeny \& Heap 1996).
This program computes strengths of absorption or emission 
components of lines in a circumstellar medium from user-input quantities such 
as disk temperature, column density, areal coverage factor (the portion of
the disk that passes in front of the star), and microturbulent velocity.
{\it Circus} combines this contribution with the {\it synspec}-computed
spectrum of the photosphere.
In the solution of the radiative transfer in its ``LTE mode," {\it circus} 
calculates the line absorption reemission along a line of sight according
to an input disk temperature, T$_{disk}$. The physical rationale for
choosing this convenient mode is that the formation of the Balmer line
flux in the disk is believed to be due to recombination, and the atomic
levels of the high-level lines and continuum may be plausibly assumed to 
be close to equilibrium with the gas kinetic temperature. 
As a reference point, we first considered temperatures near 12\,750\,K, 
the value SG01 found for T$_{disk}$ from an analysis of low-excitation 
UV metallic lines. As detailed below, at one point in our analysis we 
also used the {\it circus} ``scattering" (no reemission) approximation 
to gauge the effects of non-LTE in the disk formation of Balmer line cores.  
In addition, we point out that absorptions from the
corotating disk are formed at these same Doppler velocity as for the background 
star, i.e., no differential velocities were imposed along the line of sight. 
Although {\it circus} is capable of computing the spectrum for as many 
as three circumstellar cloud components, our initial assessment of the 
equivalent width (EW) variations suggested that the absorbing disk segments
have a complicated density stratification perpendicular to the disk plane.
Since these details can not yet be explained by {\it ab initio} models, it
was sufficient to describe the absorptions in terms of a single-component
cloud model in our {\it circus} analysis.

\section{Description of disk-induced absorptions}
\label{descrp}

\subsection{Equivalent width variations over the cycle}
\label{ewvars}

  To survey the absorptions of the high-level Balmer series in our 
spectra, we chose the members H9, H12, and H14 for EW measurement.
To these members we added the He\,$\lambda$3819 line, as it showed 
important differences with respect to the Balmer lines. 
We selected the Balmer H9 line for study because of its
proximity to the He\,I line and therefore the convenience of exploiting a 
nearly common local continuum level that would allow reliable measurements 
of line core and wing variations. The H12 and H14 lines were chosen
arbitrarily because they measure effects of yet higher atomic levels 
(such as diminished line opacity), and they 
are largely redundant to one another. This advantage allows one
to check the consistency of the EW variations.

 For our line analysis we first rectified each spectrum by identifying 
maximum fluxes along the spectrum and fitting consistent orthogonal
polynomials through these points using the {\it IRAF} task {\it continuum.}
``Raw" EWs were measured from fluxes within ${\pm 2.5}$\AA\ 
of the line centers; for H14 this window was relaxed to ${\pm 2.4}$\AA.~ 
These windows were chosen because they delimit the
absorption from the disk occultation.
For the initial measures these core depths were referenced to a pair of 
quasi-continuum points intermediate between the line to be measured and the
nearest hydrogen and helium lines. One of these points falls between the
closely occurring He\,I $\lambda$3819 and H9 lines. 
This circumstance allowed us to check 
for residual errors in local continuum placement. Such errors would produce
similarly low or high EW measures in both these lines at the same phase, 
rather than the opposing variations in these two lines caused by deviations
from the mean of the surface He/H value as the star rotates. Because
our analysis is based on absorption {\it excesses} of the line cores, 
our measurements can be tied directly to the computed absorptions 
in our disk models.
   
   We noticed that our raw EWs were significantly affected by an 
approximately single-waved variation in the line wings  that is well
correlated with the variation of the He\,I line measures and hence
instantaneous integrated He/H abundances measured by R00.
We corrected for this effect by first measuring the variable fluxes 
 in their wings at windows between ${\pm 2.5}$\AA~ and ${\pm 5}$\AA~ 
from line center. These fluxes are not affected by the disk. 
The wing EW components of the H and He\,I lines vary in perfect 
anti-correlation. We chose an arbitrary reference phase of $\phi$ = 0.35, 
which corresponds to the middle of an extended low plateau of the 
strength of helium lines (R00), signifying a near-normal He/H
abundance for the region of the star visible during this phase. 
We then corrected the core EWs using an empirical relation derived 
from the slowly varying wing and core variations for the H9 and He\,I lines 
throughout the cycle (excepting the brief occultation maxima discussed
below).  Noticing that this correction resulted in an equality of the
two H9 line EW minima, we adopted similar relations for the H12 and H14 
line EWs that forced equal EWs in their minima as well. Figure\,1 exhibits 
the corrected variations through the rotation cycle for these four lines. 
Different symbols are used for different nights' observations to best exhibit
the (generally negligible) night to night differences in our measurements at 
thoses phases observed on multiple nights.

% FIGURE 1 - 4 EW curves through the cycle
\begin{figure*}[ht!]
\centering
\includegraphics[width=12.cm,height=18.2cm,angle=90]{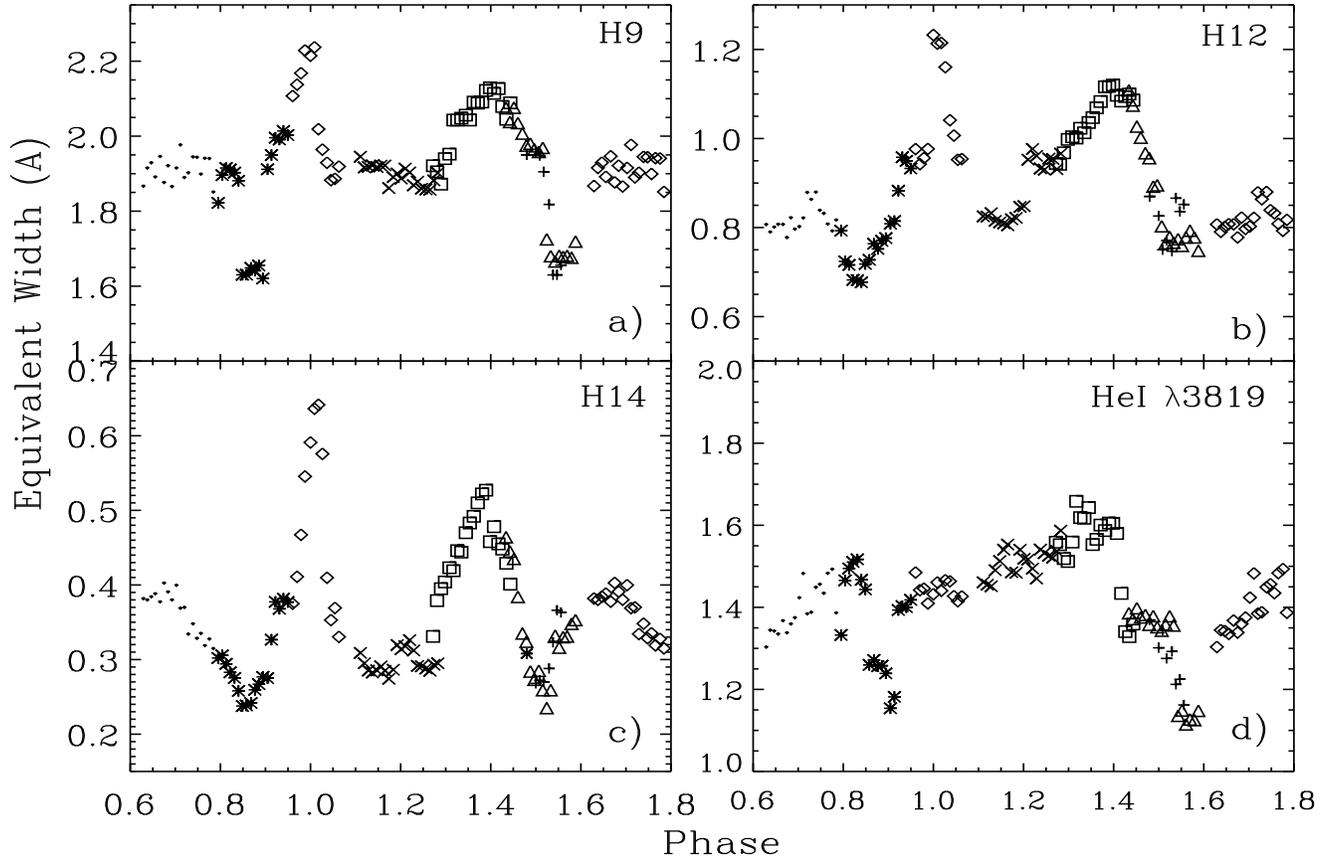}
\caption{
Equivalent width curves plotted as a function of
 rotational phase using the ephemeris in Table\,1.
The EWs are defined from local ``continuum" points located between the hydrogen
lins.  Represented are
{\it panel a:~} H9, {\it panel b:~} H12, {\it panel c:~} H14, and 
{\it panel d:~} He\,I $\lambda$3819
lines observed during 2007.  Symbols are: pluses (25 Jan), stars (29 Jan), 
diamonds and dots repeated a cycle earlier (30 Jan), triangles (31 Jan), 
squares (1 Feb), crosses (2 Feb), and diamonds (28 Feb; at $\phi$ = 0,
corresponding to observation \#800 for all lines).
}
\end{figure*}

   The panels exhibit several interesting details in the Balmer and 
He\,I line absorption curves of $\sigma$\,Ori\,E. 
First, the EW curves exhibit two maxima at $\phi$ $\approx$ 0.0 and
0.4. Distinct minima are also apparent at $\phi$ = 0.85-0.90 and 0.55. 
We will refer to these as the primary and secondary absorption maxima and
minima, respectively. The exception to this statement is that the primary
maximum is absent in the He\,I line curve and the secondary maximum is
only weakly present.  The existence of the brief minima was 
surprising because they have not been noticed in previous EW studies  
in the UV and optical (e.g., R00, SG01).  As our {\it circus} 
analysis will also show, all of the lines are optically thick, 
but the H14 line is significantly less thick than H9.  Additionally, one
notices differences in the approach of these curves to their maxima and 
minima. The edges of some of these plateaus are sharp, and this is likely 
to be a manifestation of the line opacities changing from optically
thin to thick regimes.

   The primary maximum differs in morphology from the secondary feature,
being more short-lived and triangular. Observation \#800 coincides with
the maximum EW for all the H lines.  
Following, Hesser et al. (1977), we assign $\phi$ = 0.0, 
defining the line flux minimum with the 
midpoint of this observation \#800, HJD\,2454159.687, as the epoch-defining
flux minimum. 
For completeness, we note that if one adopts Hesser et al.'s 1976 epoch
of HJD\,2442778.819, uses the cycle counts of R00, 
and assumes that the period has been constant over this 10,000 cycle 
interval, a period of 1.190841 
days would result.  However, we hesitate to claim this formally as the star's
true period (or to assign error bars) because the apparent phase difference
in photometric minimum and line strength maximum of 0.02 cycles 
suggests that caution should be exercised in
combining hetergeneous data types  - that is our $\phi$ = 0 corresponds
to 0.02 for the Stromgren filter {\it u} minimum phase of Hesser et al., 
according to the GH77 phases. If correct, this would modify the value of
the period to 1.190839 days.
We also point out that the most current
value of a wholly photometrically derived period, 1.190832 days (Townsend
2007) is only 3$\sigma$ from the above value, according to the 
errors we would perhaps simplistically assign to our own measurements.

  In addition to the maxima and minima,
minor brief changes in the absorption take place, such as that occurring at
$\phi$ = 0.80 in the strong lines H9 and He\,I $\lambda$3819. More
conspicuously, the flat plateaus at $\phi$ = 0.7$-$0.8 in the EW curves
of these lines are not shared in the weaker H12 and H14 curves, 
suggesting a thinning of absorbing circumstellar material that is 
first betrayed in the least optically thick lines. Brief minor absorption
bumps can be seen in these curves near $\phi$ = 0.20 and 0.93. 
Altogether, there is substantial evidence for the existence 
of static, high-latitude circumstellar matter. The fact that these 
absorptions are present to some extent in He\,I argues as well that 
this matter is heated more than, for example, the disk segment 
responsible for the primary absorption. The excitation of this matter 
is still influenced by a largely undiluted UV radiation field from 
the nearby star.

% FIGURE 2 - montage of diff profiles for primary max
\begin{figure*}[!ht]
\centering
\includegraphics[width=12cm,height=18.2cm,angle=90]{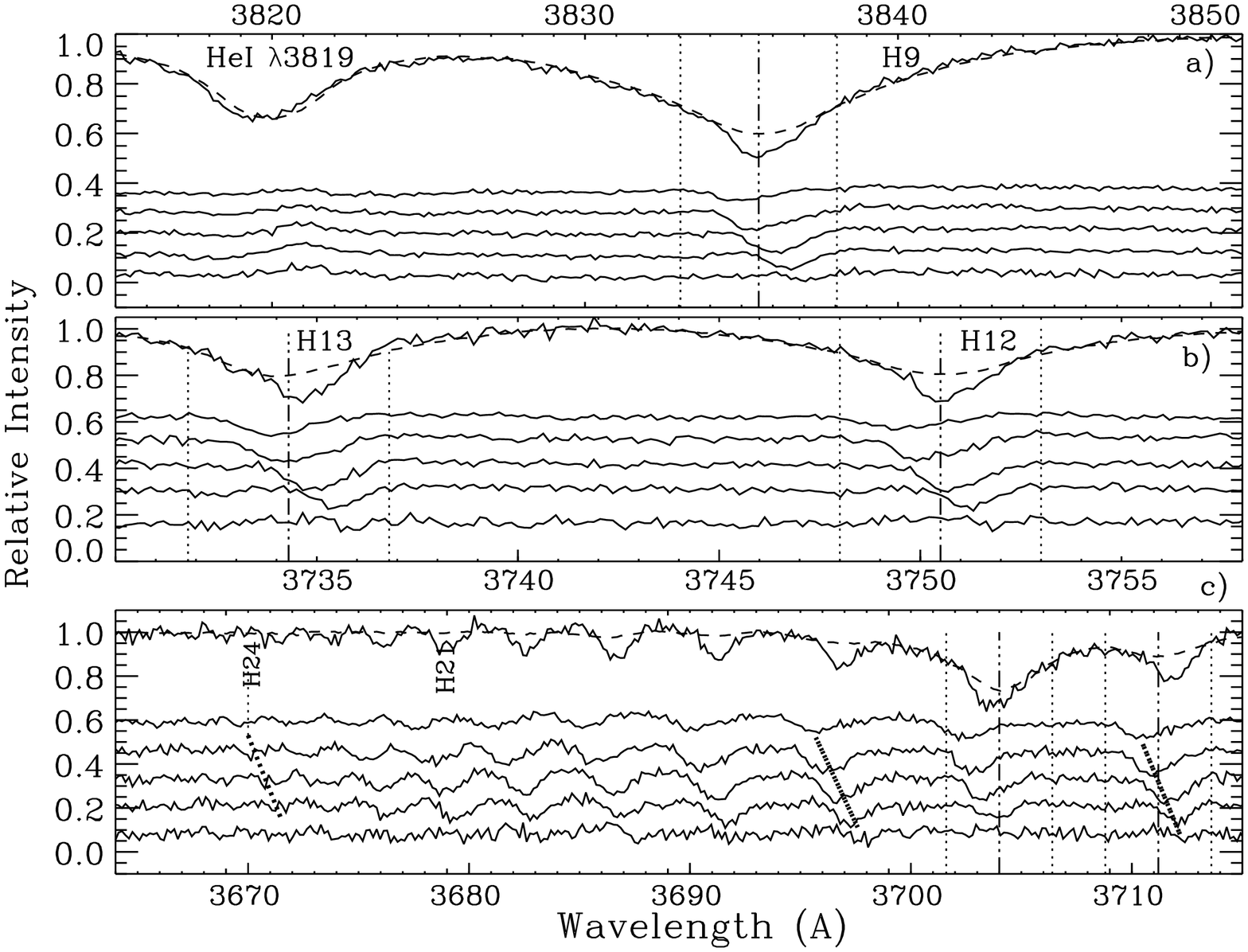}
\caption{Montage of mean and differences profiles through the primary 
occultation event. Bottom spectra depict difference
profiles with time (downward); top spectra are the mean spectrum
of the non-absorption times at $\phi$ = 0.85$-$0.90 and the observation 
\#802 ($\phi$ = 0.18) for which the central absorption lobe crosses the mean 
profile.
Panels {\it a,} {\it b,} and {\it c} represent the spectral region around
H9 and $\lambda$3819, H13 and H14, and H15$-$H24.  Sloped dotted lines trace
the migration of subfeatures in three lines with time. The difference spectra
refer to observations \#798 \& 800, 801-802, 803, 805-806, and 807-808.
These correspond to approximate phases of 0.99, 0.01,  0.03, 0.04, and 0.06,
respectively.
}
\end{figure*}

% FIGURE 3 - montage of diff profiles for primary max
\begin{figure*}
\centering
\includegraphics[width=12cm,height=18.2cm,angle=90]{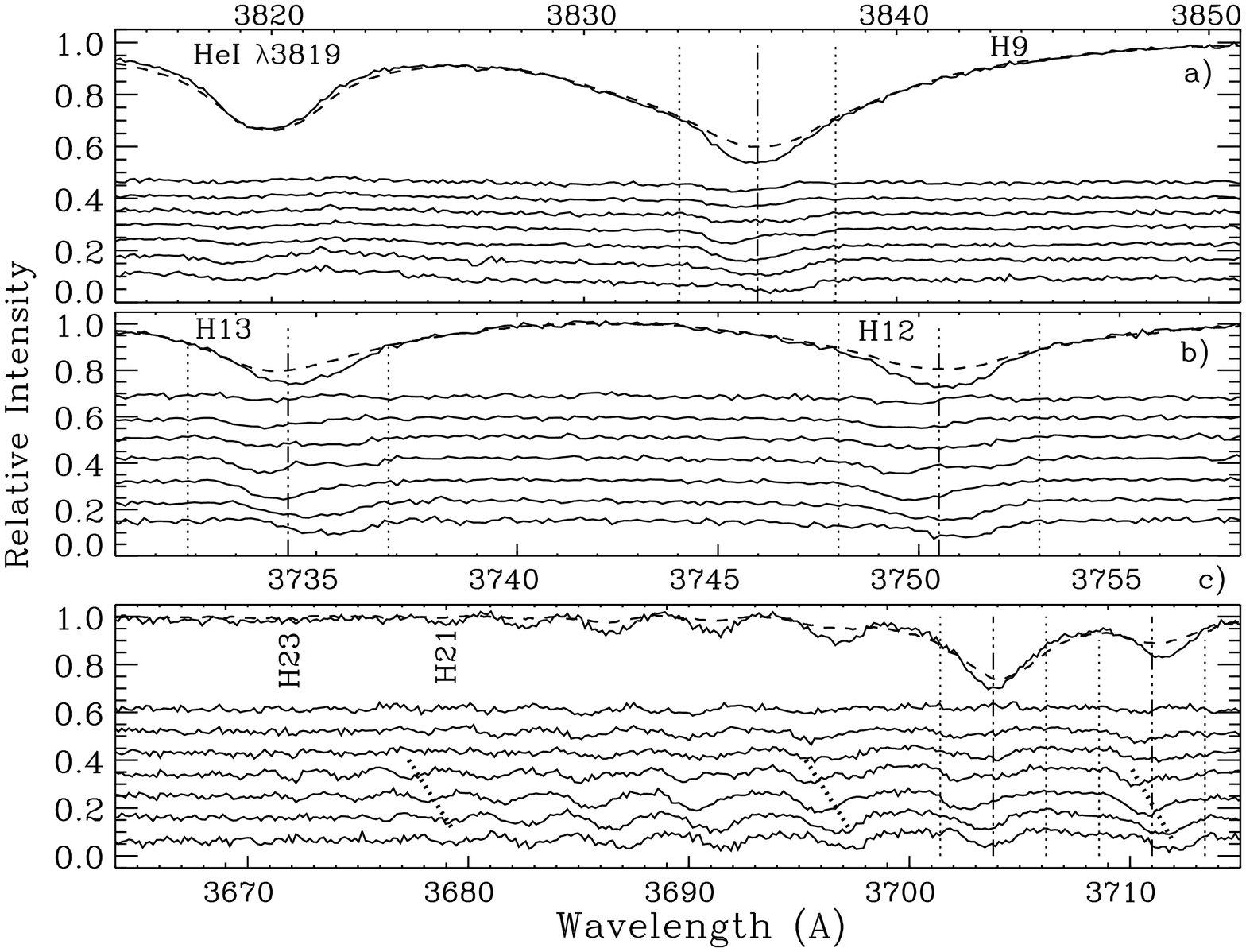}
\caption{Montage of mean and differences profiles through the secondary
occultation event. Pairs of difference spectra with time 
(downward) are depicted at bottom.  Spectra at the top are the mean spectrum
of the non-absorption times at $\phi$ $\approx$ 0.55 and the observation
\#524 for which the central absorption lobe crosses the mean profile.
Difference spectra refer to observations \# 509$-$510, 513$-$515, 516$-$518,
519$-$520, 521, 523$-$524, and 525$-$526. 
These correspond to approximate phases of .29, .32, .34, .36, .37, and .38,
and .40, respectively.}
\end{figure*}

 We end this section by discussing whether the EW variations in Fig.\,1
could be due to the passage across the disk of surface inhomogeneities, which 
could influence the line strengths either through changing the number of
absorbers in the line or indirectly through their influence on the structure
of the photosphere. To address this hypothetical problem, we stress that the
wings of these lines {\it do} respond to such changes consistently with
EW variations of the He\,I lines including the $\lambda$4713 line depicted by 
R00 (e.g.  during the phase interval $\phi$ = 0.65$-$0.80) and do not deviate 
from the anti-correlation established from this interval at intervals when the 
H\,I and HeI line cores undergo sudden changes. Second, the occultation maxima
found in Fig.\,1 develop in very short intervals, contrary to the slower
variations of at least 0.15 cycles for which the He\,I and metal line strengths
evolve (R00, Groote 2007). 

  A third reason for doubting that inhomogeneities influence our line
wing measurement derives from results using specialized model atmospheres
results appropriate to Bp stars with He and metallic patches.
In particular,  Krti\v{c}ka et al. (2007) have studied the effects of
chemical inhomogeneities on the surface of the B2p star HD\,37776, another 
B2p star with a very strong magnetic field. Both the abundance variations
across the surface and the magnetic field strength are at least 10 times 
greater than those in $\sigma$\,Ori\,E.  These
authors find that photometric variations of this star are likely due
to the effects of bound-free absorption edges produced by abundance
patches of elements like silicon (the iron abundance is depleted).
They show that the temperatures of superficial regions of the photosphere 
should be elevated by this added opacity.  Because this is the general 
region where most line cores and bound-free edges are formed, the effect 
is to cause a weakening of the line cores arising from ``passive" species,
in this case helium and hydrogen atoms. The metals and helium variations are
anti-correlated because for both $\sigma$\,Ori\,E and HD\,37776 the
metals are less depleted in belts distributed on the magnetic equator. 
Thus, the passages of patchy regions of the star's surface, irrespective of
an intervening disk, produce continuum light variations. Returning again
to our spectroscopic data for $\sigma$\,Ori\,E, and utilizing
the Krti\v{c}ka et al. simulations, the cores of hydrogen and helium 
lines should weaken during these phases -- yet they do not.
We note in particular that this fact shows that the slight enhancement of
the He\,I $\lambda$3819 line in Fig.\,1 cannot be attributed to surface
inhomogeneities on the star.

\subsection{Line profile variations}

  To represent the line profile variations of the two occultations, 
we compared individual or pairs of line profiles differenced against
average profiles of the two minima. Specifically, for the first stretch of
absorption minimum centered at $\phi$ = 0.87
the mean was computed from 
observations \#305$-$311, and for the second minimum ($\phi$ = 0.55)
from observations \#448$-$457. Figures\,2 and 3 exhibit, at the bottom
of each panel, the migration of absorption lobes across 
the profile for He\,I and H9, H12 and H13 line, and H15 and higher
for both occultation events.  
At the top of each panel we depict the rectified unabsorbed mean spectrum 
and an example of one of the observations during the occultation. 
Close inspection of the difference spectra during the second occultation
discloses the passage of two lobes across the profile. 
In our depiction 
the two lobes can be discerned in Fig.\,3 by the flat-bottomed difference
profiles of observations \#513-515 ($\phi$ = 0.31-0.33), followed by a 
second flat-bottom in the \#523-524 ($\phi$ = 0.38-0.39) profiles.
The second lobe in the second occultation
can be parameterized by a $\sim$50 km\,s$^{-1}$ gaussian and is deeper 
than the first one. The first lobe of this occultation, as well 
as the lobe of the first occultation, are slightly narrower and
can be fit with a $\sim$40 km\,s$^{-1}$ gaussian.

   From Fig.\,2 we estimate that the midpoint of the first occultation,
as judged from the transit across the profile of the centroid of the 
excess absorption lobe,
occurs for observation \#802 ($\phi$ = 0.018), or $\approx$0.02 cycles later 
than the observation responsible for the maximum hydrogen EWs in Fig.\,1. 
Our error estimate given above was obtained from the 0.02  
cycle difference between the light minima and line profile passage events
found by GH76.

  The difference between times of passage of the absorption line in the
profile and EW maximum for the whole line carries over to the
second occultation. From Fig.\,3 because of the finite signal to noise,
the breadth and unequal depths of the features we cannot establish a
reliable phase for the mid-point of the occultation. Our best estimate 
is at the beginning of observation \#524 ($\phi$ = 0.390), 
which would be some 0.37 cycles 
after the central lobe passage in the primary event and identified with 
observation \#802 ($\phi$ = 0.018). 
Thus, whereas there is a separation of 0.40 cycles
between the EW maxima, the separation between the lobe passage events
is somewhat smaller.
 Similar structure can be discerned in the semi-resolved double
lobes, centered at $\phi$ $\approx$ 0.35 and 0.42 of the secondary 
occultation in the light curves of
Hesser et al. (1977) and Oksala \& Townsend (2007).
Perhaps related to the complicated shape of the secondary EW maximum
is the fact that our best estimate of the separation between this
and the primary occultation, 0.40 cycles, is about 0.02 cycles smaller
than the value determined from the Hesser et al. (1977) and the nearly
contemporaneous Oksala \& Townsend (2007) light curves. 
It is not clear to us whether this minor discrepancy has any 
consequences on the details of the disk model. The most conservative
assumption is that the opacity (column density) of the disk is unequal
at opposing positions on either side of the central plane. 
As already suggested from resonance line data (GH97, SG01), a 
disk ``warping" could explain in principle both the appearance of a minor
lobe within the secondary occultation event as well as an {\it apparent}
unequal fall off in column density at small and opposite phases from
the central maximum absorption. This circumstance would likely affect
optically thin (photometric) and thick (Balmer line) measurements
differently.

In our interpretation, this implies asymmetric
density distributions above and below the magnetic planes; that is, this
disk segment has partially separated layers.

   In the curves of the primary and secondary occultations (Figs.\,2c 
and 3c) the Balmer line series can be discerned out to H24 and H23,
respectively. (We note that differences formed against the mean of all
spectra except those in the two occultations have a higher signal quality 
and bring out these limiting lines more clearly.) We will return to these
limiting line members in our later discussion of electron densities in
$\S$\ref{fttng}. In the lower panels of Fig.\,2 and 3 we have also indicated the
acceleration of the absorption lobes through the profile. A 
cross-correlation analysis of the shifts of the line with respect to a
fiducial spectrum discloses that the average acceleration of these features 
is 72 km\,s$^{-1}$\,hr$^{-1}$ and 50 km\,s$^{-1}$\,hr$^{-1}$ for the
primary and secondary occultations, respectively. The larger accelerations
of the features occurring during the primary occultation (assuming
the disk is rigidly rotating) indicate that it lies 
further from the star than the disk segment producing the secondary
occultation. As pointed out earlier, the absence of the He\,I line absorption 
during the primary occultation is likewise consistent with this segment 
lying further from the star.

\subsubsection{Balmer line and continuum absorptions }

   In preparation for a {\it circus} analysis of the physical parameters 
of the absorbing disk segments, we have computed ratioed spectra of the
means of the raw (unrectified) spectra. These exhibit the largest absorptions 
during the two occultations (observation \#800 for the primary, and observations
\#516-524 ($\phi$ $\approx$ 0.36) for the secondary) with respect to the 
absorption minima spectra 
occurring at nearby phases (observations \#305$-$311 and observations 
\#448$-$457,or $\phi$ $\approx$ 0.86 and 0.56, 
respectively).  These spectra and the ratios are depicted in the two panels 
of Figure\,4. These panels show the clear absorptions in both the lines and 
the effective Balmer continuum, which starts at $\lambda$3690$-$3700. 
The ratios $F(\lambda3650)/F(> \lambda3700)$ for the occultations come to 
about 0.845 and 0.925. We will use {\it circus} to reproduce these ratios 
in the next section.

% FIGURE 4 - comparison of pure star and occultation spectra (2)
\begin{figure}
\centering
 \includegraphics[width=6.5cm,height=9.8cm,angle=90]{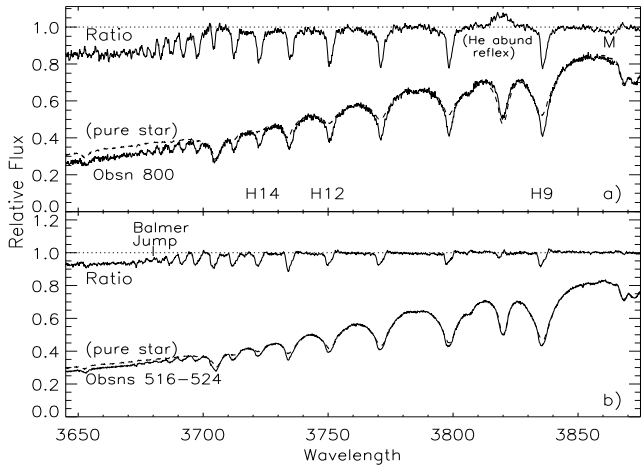}
\caption{
  The spectrum exhibiting the maximum absorption during
central occultation and the mean spectrum representing no disk absorption
referenced in previous figures. The ratio of these spectra are depicted
at the top. Panels {\it a} and {\it b} represent the primary and secondary
occultations, respectively. In {\it a}, by ``M" we denote metallic line features
used in the text to set a rough disk temperature and also the weakened
helium line (shown as apparent ``emission") due to chemical inhomogeneities.
The Balmer jump is represented by the lower flux shortward of about
$\lambda$3690.
}
\end{figure}

  It is worth pointing out that the Balmer continuum absorption in 
these particular pairs of specra are
unlikely to be contaminated significantly by the effect of telluric 
UV absorption through different air masses.
For example, in the case of the primary occultation, the observation
was obtained at an hour angle (HA) of 0:49, as compared to the nearly 
equivalent value of a mean HA of 0:47 for the observations comprising
the minimum spectrum. The respective HAs for the secondary event are
likewise practically identical: 0:33 and 0:36. These differences
are less than the exposure times of the spectra (15 minutes).

  The hydrogen line absorptions show no noticeable radial velocity shift
($< \pm{5}$ km\,s$^{-1}$), and they are confined entirely within 
${\pm 2.5}$\AA\ of the line cores.
One anomaly in the primary occultation spectrum (Fig.\,4a) is that
the He\,I $\lambda$3819 actually appears {\it weaker} during the
occultation. This effect reflects the passage of a He-rich spot at
phases $\phi$ = 0.8--0.9 (R00), from which spectra
used to compute the {\it hydrogen} line minimum were taken.

\section{ Determination of disk parameters}
\label{fttng} 

   The primary estimates we are able to make for the disk segments causing
the hydrogenic absorptions come from analysis with the {\it synspec} 
and {\it circus} models.
We adopted a solar H/He abundance in our disk models.
However, since these are unlikely to be strictly correct (assuming the He/H
abundance in the disk sustantially reflects the values for the material at both
magnetic poles), the column densities we derive should be understood to 
refer to hydrogen atoms.

\subsection{Rotational and magnetic axes inclinations}
\label{angls}

  We have discussed the likelihood already from the discussion of 
EW and line profiles that a pair of disk segments
residing near the intersections of the rotational and magnetic equators 
occult different regions of the star at times separated by approximately one-half of the
rotation period.

  Our disk analysis begins with an estimate of the stellar properties.   
We used Kurucz (1993) models with parameters T$_{\rm eff}$ = 23\,000\,K 
and log\,g = 4 (e.g., Smith \& Groote 2001).  We computed the 
photospheric line profiles with a rotational velocity of 150 km\,s$^{-1}$.
The foregoing geometric properties of the disk are not sensitive to these
estimates.

    The geometric constraints begin with a determination of 
the inclination angle $i$ of $\sigma$\,Ori\,E's rotational 
axis as well as the angle between the rotational and magnetic axes 
-- the so-called magnetic obliquity, $\beta$. 
Traditionally the former comes from comparing the star's projected
rotational velocity with its radius as inferred from its position on
a stellar evolutionary track. Estimates in the literature range 
from $i \approx 90^{o}$ (Bohlender et al. 1987, 
Short \& Bolton 1994, TOG05), to $i \approx 39^o$ (R00).
The latter value is based on an uncertain distance estimate. 
From detailed fitting of the light curve
of the star to their Rigidly Rotating Magnetosphere (RRM) model, TOG05
found that $\beta + i \approx 130^{o}$. 
They adopted $i \approx 75^o$ and $\beta \approx 55^o$. 
The intermediate value of $\beta$ is based largely on the unequal spacings 
(0.4, 0.6 cycles) of the light minima of the light curve,
which implies that the center of the stellar magnetic field is
displaced 0.3R$_{*}$ perpendicular to both the rotational and magnetic
axis and is also displaced 0.05R$_{*}$ along the magnetic axis. We note
also that Groote (2003) gives $\beta$ = 67$^o$. However, this value is based
on an analysis of the distribution of He/H on the surface. 
In particular, it is likely that the centroid of the He-rich patch is 
somewhat displaced from one of the two magnetic poles (e.g., Groote \& 
Hunger 1997).
An additional set of constraints can be placed on these angles 
from the shapes of the equivalent profiles of the two maxima. As already 
mentioned, the primary maximum in Fig.\,1 has a triangular shape, and the
two legs of this triangle have unequal durations. This occultation appears
to be the one for which the segment passes most closely to the projected center
of the star in the observer's line of sight. At $\phi = 0.4$ the opposing
segment occults only the lowermost limb of the star during mid-passage.
We will now investigate the shapes and intensities of the occultation
profiles.

  We have designed and discussed a disk occultation tool to quantify the 
absorption of light and equivalent width curves from an optically thick disk 
or, alternatively, from a thick disk surrounded by a translucent periphery 
at greater distances from the magnetic plane (see discussion and Figs\,6-7 
in Smith et al. 2006). 
We found that a triangular shaped occultation profile can be reproduced if the
disk has a limited radial extent (i.e., it is ring-like) and/or its height
perpendicular to the magnetic plane, is small.
Another signature of a ring-like structure is if the ingress/egress
edges of the occultation event are sharp. This can happen only if the
disk is separated from the star.

    Additional considerations in evaluating the disk geometry in our 
model are first, that the ratio of secondary to primary line and 
continuum absorptions is $1.3-1.4$. We will use this range as a starting
guess for the ratio of the stellar areas eclipsed by the disk segments. 
The amounts of the absorptions in the {\it circus} computations 
discussed below will also translate to the individual occultation areas.
This leads to a secondary constraint that the angles $i$ 
and $\beta$ must be near 90$^o$ and not much larger than 55$^o$, respectively. 
Otherwise, the measured Balmer continuum absorption cannot be attained 
by an occultation that is too weak when the disk crosses the ``lower" limb 
of the star, as seen from our vantage point.
Finally, the fact that there is just a hint of an M-like reversion in the
secondary occultation profile limits the value of $i$ to at least than 
$80^o\pm{5}^{o}$. (However, if the two peaks result from disk warping this
argument is questionable.)
With the constraints we imposed above as well as TO05's arguments,
this fixes the $\beta$ value as near $55^{o}\pm{5}^{o}$. These values are 
in reasonable agreement with estimates based on different criteria by
other authors. 

  For completeness, we point out that although the double lobes found during 
the secondary maximum are not treated explicitly in the foregoing analysis, 
it is clear that this disk segment is double-layered such that the density in
or very near the magnetic plane is lower than in the two layers just outside 
it.  We also note that for $i \ge 80^o$, a period of 1.19 days, and its 
$v \sin{i}$ of $140$-$162$ km\,s$^{-1}$ (e.g, Bolton et al. 1986, R00,
the mean radius of $\sigma$\,Ori\,E comes to $3.4-3.9$\,R$_{\odot}$.
As is by now well known (e.g., Hunger, Heber, \& Groote 1989), 
these radius estimates seem low for a B2 main sequence star, 
but we can offer no solution to this apparent dilemma.

\subsection{Disk dimensions}

  The dimensions of the disk segment responsible for the primary occultation 
can also be inferred by appealing to H$\alpha$ emission studies, which provide
estimates of $5-6.4$\,R$_*$ for the outer disk dimension (Groote \& Hunger
1982, Short \& Bolton 1994, TO05), the accelerations and durations of the 
absorption lobes in Figs.\,2 and 3, and the areal coverage arguments of the 
previous section.  Our simulations also show that for the primary occultaion
the outer radius must be at least 4\,R$_*$ in order for the disk to have a 
finite thickness in the radial extent.  
Table\,2 gives the dimensions of the absorbing
disk segments in units of stellar radii as well as the mean densities and
temperatures of our solutions. The radial thickness refers to the quantity
R$_{outer}$ - R$_{inner}$.  We note that the derived radius
is consistent with that given from the durations of the lobe apparitions
found in  Figs.\,2 and 3. The errors for these parameters are about
${\pm 20}$\%.
For the primary absorption the results pertain to the rise time phase
for the first half of the event, which lasts 0.04 cycles. The shorter span
for the decline
(0.03 cycles) of the second half can best be simulated by decreasing the 
semiheight from 0.3R$_*$ to 0.2R$_*$. These parameters also preserve the 
triangular shape of this primary occultation profile.

   The challenge for deriving the dimensions for the disk segment causing the
secondary absorption maximum is that they should be large enough to ensure 
enough areal coverage of the star and to produce the observed absorption in the
Balmer lines and the Balmer jump. In general, lower values of $\beta$ and 
R$_{outer}$ and larger $h$ values insure this. 
A value of R$_{inner}$ $>$ 2R$_{*}$ is impossible to reconcile with the 
observed amplitude and shape of the occultation profile with phase in our 
models with the assumed $\beta$ and $i$ angles.  
However, if one adopts the view that the extended lifetime of the
occultation is due to seeing two different radial zones of the same segment
along a line of sight (i.e., the disk is warped), one can relax the distance
R$_{outer}$ up to 4R$_{*}$ (but no higher).  Since this constraint
is consistent with the ratio of the radii suggested by the two acceleration
rates determined in $\S$3.2, we cannot rule this out. We have added this 
possibility in Table\,2 by showing the admissible 
ranges of the inner and outer radii.
parameter to the assumption of disk warping, i.e. 2R$_{*}$ $\le$ R$_{outer}$
We also point out that in the warping interpretation, R$_{inner}$ can move
in to 1.2R$_{*}$, so the disk extends {\it almost} to the star's surface.

  According to our picture, at the middle of the first occultation at 
$\phi$ = 0 the whole of the disk segment is silhouetted against the star.
For this portion of the disk we may estimate the mass from the dimensions
of the segment (area and column density) and its density. This gives
a mass of about 5$\times$10$^{-11}$ M$_{\sun}$. If we take half the
mass loss rate as the rate of resupply of this hemispheric disk segment 
(1.2$\times$10$^{-9}$ M$_{\sun}$\,yr$^{-1}$; Krti\v{c}ka, Kub\'at, \&
Groote 2006), the mass loss will replenish this disk segment in about 2 weeks. 
Because a small fraction of the wind emanating from close to the magnetic 
poles will avoid being focused into the disk and, further, because we may 
not see the full disk segment in azimuth around the hemisphere, the actual 
lifetime of the particles in the disk will be somewhat 
longer.\footnote{The fact that our column densities are 
averaged over the disk height will cause the total density and hence mass 
in our model to be underestimated. We may reasonably expect the longevity 
of particles in the disk to be a few months.} Note that the transit of a 
disk across the star lasts only $\ltsim$10$^{-3}$ of this time. 
Therefore, leakage of disk particles either into the star or through its 
outer edge would produce a tapered absorption wing that is too small 
to be visible in our data.

  We conclude this discussion by  calling attention to the 
EW minima within the general model for the disk, which occurred on a total
of three of the seven nights of our observing run on this star.
One explanation for these minima is that a second translucent
component exists layered over the more confined opaque disk. In this 
explanation, we would be looking through a light ``haze" (with about 1\% of
the column density) through most viewing aspects.
However, little if any tail is observed in the curves of Fig.\,1. (This fact 
underscores our finding above that the density contrast of the disk proper
relative to circumstellar matter at moderate or high magnetic latitudes is
rather high.)  A second possible way to explain the EW  minima is that 
they are ``reflexes" of small incipient emissions that may still be 
detectable in the high level lines. Indeed, FEROS observations of the H$\alpha$
line at these phases (Groote 2007) exhibit strong emissions and an elevation
of the line core at these phases.  In this event there would be no need of a 
pervasive low-absorption component.

\begin{table*}[ht!]
\begin{center}
\caption{\centerline{Derived absorbing disk parameters 
(assuming $i$ = 80$^o$, $\beta$ = 55$^o$)}
}
\centerline{~}
\begin{tabular}{lr|lr}  \hline\hline
Primary Occultation: &   &  Secondary Occultation: &  \\
 &  &  & \\
Outer radius &   $5-6{\pm 0.5}$R$_*$ & Outer radius (2 solutions)  &    $2-4R_*$ \\
Radial thickness  &  $2.5-3R_*$ & Radial thickness   & $0.25-2.8R_*$ \\
Height   &  0.3${\pm 0.07}$R$_*$ & Height    &     0.45${\pm 0.10}$R$_*$  \\
Temperature & 10,500K ${\pm 1000}$K & Temperature  & 12,000K  ${\pm 1000}$K \\ 
N$_e$            &   1${\pm 0.5}$$\times 10^{12}$ cm$^{-3}$ & N$_{e}$           &    1${\pm 0.5}$$\times 10^{12}$ cm$^{-3}$  \\
Column density &   1${\pm 0.5}$ $\times 10^{24}$ cm$^{-2}$  & Column density & 1${\pm 0.5}$$\times 10^{24}$ cm$^{-2}$  \\
Areal coverage &  0.45${\pm 0.10}$  & Areal coverage &   0.31${\pm 0.10}$  \\

\hline
\end{tabular}
\end{center}
\end{table*}

\subsection{Thermodynamic parameters of the disk segments}

  The Balmer absorption spectra of Fig.\,4 provide sufficient information 
to estimate the temperature, electron density and column density of the 
intervening disk segments. To make these estimates, we must assess the
importance of errors in various factors used in these calculations.
First, and importantly, the assumed areal coverages depend strongly
on the geometrical angles derived above. Despite our agreement with
earlier authors, errors in our estimates of these parameters propagate
to the areas in complex ways and should be considered the largest source 
of our uncertainty.  Second, it turns out that the disk microturbulence,
$\xi$, is {\it not} an important factor. We find almost identical solutions 
for fitting excess Balmer line and continuum absorptions
for values of $\xi$ in the range $0-20$ km\,s$^{-1}$. 
We have arbitrarily adopted a value $\xi$ = 10 km\,s$^{-1}$. Third, we have  
used non-LTE model atmospheres computed from the BSTAR2006 grid (Lanz \& 
Hubeny 2007) and found that the photospheric Balmer jumps for LTE and 
non-LTE model atmospheres differ negligibly 
given the adopted T$_{\rm eff}$ for $\sigma$\,Ori\,E.
This can be inferred also by examining Figure\,6 of Lanz \& 
Hubeny. These authors find that changes in abundances have little effect
on photospheric temperatures in the range $\tau_c = 0.01 - .003$, where 
much of the Balmer line cores is formed.

  SG01 pointed out that the presence of certain low resolution line 
aggregates in the far-UV spectra suggest the presence of a cool 
component to the disk of $\sigma$\,Ori\,E having a temperature 
of about 12\,750\,K\,$\pm\,{750}$\,K.\footnote{ For studies of different
lines and in a different wavelength regime, we estimate that the 
uncertainty of this mean disk temperature should be at least doubled.
This uncertainty does not propagate to the more precise temperature
{\it difference} quoted below for opposing disk segments. } 
In this study we discovered
a weak but important diagnostic in the excess absorption spectrum of the
secondary occultation (Fig.\,4b), namely weak absorption blends at 
$\lambda$3760 and $\lambda\lambda$3855$-$60 arising primarily from transitions 
of once ionized light and iron-group metals. Our {\it circus} syntheses
show that these lines appear for T$_{disk}$ $\le$12\,000\,K over disk 
column densities of interest. 

  Our primary modeling diagnostic for column density and areal coverage
is the excess Balmer jump noted in Fig.\,4. While potentially
useful, the Balmer lines themselves were found to be enhanced primarily
by non-LTE effects. Thus, whenever the excess absorption below 
$\approx \lambda3647$ could be fit, the disk Balmer line spectrum is
far too strong. We will therefore confine ourselves to fitting the Balmer 
jump in the following discussion.

  The electron density N$_e$ of the disk medium can be determined
by the presence of the highest-level Balmer lines in the occultation
spectra (Fig.\,4). One might first draw an analogy to the prominence of
these lines in the spectra of B supergiants. For low density atmospheres,
the lines are visible because the weak Stark wings cause less blending.
This motivated us to inquire whether the Inglis-Teller (1939) criterion 
might apply in this instance. This condition gives  as an estimator 
that the last visible line of an hydrogenic spectrum scales with the
electron density as N$_{e}^{-7.5}$. By comparing
the highest visible Balmer lines in the spectra of early-BIa stars
(e.g., Bagnulo et al. 2006) and appealing to model atmospheres for stars
of these temperatures and gravities (e.g., Crowther \& Lennon 2006), 
we found that the visibility of H24 and H23 
in occultation spectra suggest N$_{e}$ values of 1$\times 10^{12}$ and 
2$\times 10^{12}$ cm$^{-3}$, respectively. However, this argument may not
be precise as our spectra show that the strengthening of these 
lines occurs mainly in the line cores. 
Our modeling of these absorptions 
suggests that what is important in determining the visibility of the last 
Balmer line is the column density of high-level hydrogenic atoms producing 
the absorption.
In particular, we find that through the Saha equation N$_{e}$ is largely
determined by fixing the column density at which the optical thickness 
of a high-level line is $\approx$1.  For temperatures of interest 
(10\,000--14\,000\,K) the density corresponding to the semi-visibility of 
H23--H24 is also N$_{e} = 1\times 10^{12}$ cm$^{-2}$.
We found that the fitting of the Balmer continuum absorption is insensitive
to the value of N$_{e}$. Therefore, we adopt this density for both disk
segments in our {\it circus} analysis of the disk segment parameters. 

   We adopted the densities from the above arguments to determine
disk temperatures, column densities and fractional areal coverage 
factors needed from the Balmer jump found below $\lambda$3700 in Fig.\,4. 
To add to these constraints, we also invoked the condition that the areal 
coverage ratio should be $1.3-1.4$.  
However, we modified this by determining that a 
correction of about 10\% for limb darkening (assuming a coefficient $\mu = 
0.6$) is necessary to account for the transit of the disk segments across 
noncentral regions of the star. This consideration brings us to a ratio of 1.5 
for the fractional areas. This condition is met by the entries in Table\,2. 
Our models established quickly that the hydrogen column density should 
be near 1$\times$10$^{24}$ cm$^{-2}$. 
(However, the column lengths are not determined well enough to compare 
the radial distances of the two disk segments.)
This condition derives from the 
broad opacity plateau at $\approx$ 1$\times$10$^{24}$ cm$^{-2}$ in this 
wavelength region for the relevant temperatures. 
For substantially larger column densities the disk plasma becomes optically 
thick in the continuum, 
and the Balmer jump it forms begins to decrease. For smaller column
densities one runs quickly into unrealistically large fractional areal
coverages and the Balmer jump decreases owing to optical thinness of
the Balmer continuum. Given these considerations, we believe that 
the column densities we found, which are coincidentally the same
for the disk segments causing the two occultations, are accurate to
within a factor of two. The fractional coverages we infer depend on
this condition combined with the absorption ratio of $1.3-1.4$ derived
above from the two occultations.
  
   Our estimated column density agrees well with the derived values
of the electron density, 10$^{12}$ cm$^{-3}$, the inferred radial extent
of the disk (2.5$-$3\,R$_*$) for the segment causing 
the primary maximum and within a factor of two of that for the segment
producing the secondary. 
Note that SG01 found a column density of 3$\times$10$^{23}$ cm$^{-2}$ 
from the strengths of  
absorption of Fe aggregates and low excitation light-metal lines in the UV.
When account is taken that SG01 assumed for simplicity a full occultation
of the stellar disk, their column density is virtually the same as the
value we derived from the assumed density and the values 
given in Table\,2, 3$-$7$\times$10$^{23}$ cm$^{-2}$.
Our solution, assuming $\tau_{continuum}$ $\approx$ 2 and disk homogeneity, 
likewise agrees perhaps fortuitously well with the peak value 
$\tau_{continuum}$ $\approx$1.6
determined by TOG05 for the optical thickness viewed along the central
magnetic plane. The latter determination was made for a decentered
dipole that probably most accurately reflects the magnetic geometry.
  
  Our homogeneous disk solution overlooks the evidence that the segment
producing the secondary occultation occurs in two semi-resolved stages.
As already mentioned, one interpretation of this is that the a warping
occurs in the inner disk, as suggested by previous studies. In any case,
this semi-resolved structure of the occultation is 
noticeable in light curves at several wavelengths as well.
It may be that what observers identify as ``warping" is in 
fact due to a departure from strict coplanarity predicted for extensive Bp 
disks for which the magnetic $\beta$ obliquity has an intermediate value 
(TO05).

  Returning to the Balmer lines, we were able to 
evaluate departures from LTE in the line source functions
by comparing the column densities needed to fit the Balmer
jump with those in the lines. As already stipulated, we assumed that the 
Balmer continuum flux is formed in the disk in near LTE. By comparing the 
column densities needed to fit the lines to the Balmer continua, we found
for both occultations the same approximate ratios of source to Planck
functions: $S/B$ = 0.006 for
H9 and $S/B$ = 0.1 for H20.

   Finally, the temperatures for the two disk segments are consistent with 
the radiative temperatures one would expect for radiatively excited matter 
close to the star. There is no evidence in these disk segments in particular
for the heating responsible for the presence of the variable N\,V resonance 
lines (SG01).

\section{Summary and Conclusions} 

   We have used high-level Balmer absorption data finely sampled around the 
rotation cycle to show that the distribution of plasma in the magnetosphere 
of $\sigma$\,Ori\,E is more complicated than previously believed. 
Most especially, the secondary occultation is comprised of two semi-resolved 
events.  In addition, there seem to be 
at least two brief appearances of absorbing matter well
beyond the disk plane.
The disk plane is dominated by two
segments which occult portions of the star some 0.4 cycles apart. 
We have no direct information from absorption profiles alone on the 
distribution of matter at other azimuths in the plane and so look
forward to future analyses based on the H$\alpha$ and/or H$\beta$
emissions over the cycle. 

 The occulting disk segments are sections of a ring which occult the
star as the frozen disk corotates around the star.
The segment causing the primary occultation lies further from the star 
(between limits of $\approx$2.5R$_{*}$ and 6R$_{*}$) than the opposing segment. 
There are at least three arguments
that point to the disk segments lying at different orbital radii. First, 
gas in the segment causing the primary occultation has a lower temperatures
(10\,500\,K, compared to 12\,000\,K). This can be inferred first from the 
lack of an enhanced He\,I feature and also from the absence of weak absorptions
of metallic-line
blends referred to in $\S$4.3. Second, the acceleration of spectral
migrating components across the line profile is higher for this segment.
Third, our occultation analysis program indicates that one cannot reconcile
the radii of the two segments even if one attributes the broad secondary
absorption maximum as being due to viewing two occultations of a single
warped disk segment.

  Perpendicular to the magnetic disk, the semiheights of the two segments
are some 0.3$-$0.45R$_{*}$. 
It is important to stress that the disk of $\sigma$\,Ori\,E 
has a smaller semiheight than disks for other disk-harboring Bp stars 
that we have analyzed with similar modeling techniques
(e.g., SG01, Smith et al. 2006). It is tempting to speculate
that a disk should be more compressed toward the plane due to the
pressure provided by a large magnetic field (e.g., Babel \& Montmerle 1997).
However, this cannot be the full story because the most relevant quantity, 
the ratio ``$\eta$" of magnetic to wind energy densities
is actually larger for 36\,Lyn than for $\sigma$\,Ori\,E and yet 
the disk of the former has a somewhat larger height
(Smith et al. 2006). Might other considerations, such as the larger
radial extent of the disk (e.g. in the case of 36\,Lyn), 
have a bearing on this question?

   In addition to these dimensional properties, we find that  a
phase discrepancy of +0.018 cycles exists between the phase at which the
central absorption transits the center of the hydrogen line and the phase of 
EW maximum. This agrees with a similar phase difference found by GH77 and 
calls into definition of the star's period when determined from different
observation modes. The secondary occultation is distinguished by the passage
of two weak lobes, the phases of which are consistent with the semi-resolved
structure of the light curves in the literature. The interpretation of this
structure is not clear, but it may be related to a disk ``warping" or a
corrugation of the disk plane predicted in the TO05 models.

  When we view the disk edge-on through the magnetic plane, the 
Balmer line absorptions are opaque, and thus the absorptions of the high-level 
members of this series increase.  For H9 the optical depth 
is about 100, while it is $\approx 10$ for H20, and $\approx 2$ at the Balmer 
edge.
As the disk ingresses and egresses there is an absence of an absorption tail
in the equivalent width curves. 
This confirms the theoretical expectation
(e.g., TO05) that the density scale height is very small compared to the
stellar radius.  This implies in turn that our homogeneous slab models
of the disk and the total mass estimate of about 1$\times$10$^{-10}$
M$_{\sun}$ are only rough approximations. 
 
   How would hypothetical data of indefinitely high quality (signal-to-noise, 
spectral resolution, and cadence) improve on our analysis of the disk of 
$\sigma$\,Ori E? We can think of the following applications of 
enhanced quality data:

One improvement would be in the mapping of vertical density distribution 
and indeed the separation of the two density maxima associated with the 
secondary occultation. This would result from tracing details of the 
absorption subfeatures as they migrate across the line profiles. An
improvement in our understanding of the vertical stratification should
indirectly improve out understanding of whether the disk is warped and
the inner edge of the secondary occultation-causing segment extends almost
to the star or not.
It will also clarify the cause of the phase delays that both we and GH76 
have found among various hydrogen lines and with respect to the continuum. 
A resolution of this issue would allow an improved determination
of ephemerides from flux curves derived from different diagnostics,
thereby leading to a definitive determination of the period and perhaps
its first derivative.

 Second, the changes in the highest-level lines detected during the
ingress and egress of the occultations can be used in principle to
further map the vertical density distribution of the disk segments
because of the different optical depth regimes in which the various
lines are formed.

  Third, highly accurate difference profiles should allow one to
search for asymmetric absorption features in the EW curves that could 
indicate the rate at which disk particles leak from the disk either 
through the inner or outer edge.

  Fourth, increased data quality improvements (assuming a stable 
spectrograph and atmospheric transmission) can lead to the observation of
subtle changes in disk structure. We suggest that these might 
occur on timescales of one to a few months or longer.  Finding such 
changes could allow one ultimately to correlate them with the expected
``break out" of disk matter through the outer edge (ud-Doula, Townsend, 
\& Owocki 2006), possibly associated with X-ray flares from this star,
(Groote \& Schmitt 2004, Sanz-Forcada, Franciosini, \& Pallavicini 2004). 

\begin{acknowledgements}
The authors are indebted to Drs. Detlef Groote and Rich Townsend 
for key discussions on phase discrepancies and surface inhomogeneities 
that improved the quality of this work.
DAB also thanks Dr. Dmitry Monin for conducting the critical primary occultation phase
observations of $\sigma$\,Ori\,E on the final night of the observing run.
\end{acknowledgements}


\begin{thebibliography}{}

\bibitem[]{} Babel, J. \& Montmerle, T. 1997, A\&A, 323, 121

\bibitem[]{} Bagnulo, S., Cabanac, R., Jehin, E., et al. 2006, The UVES
Paranal Observatory Project, 
http::\/\/www.sc.eso.org\/santiago\/uvespop\/field\_stars\_uptonow.html

\bibitem[]{} Berger, J. 1956, Contr. Inst. Ap. Paris, Ser. A, 217

\bibitem[]{} Bohlender, D. A., Brown, D. N., Landstreet, J. D., et al.
1987, ApJ, 323, 325

\bibitem[]{} Bolton, C. T., Fullerton, A. W., Bohlender, D. A., et al. 1986,
The Physics of Be Stars, ed. A. Slettebak \& T. P. Snow (Cambridge: Cambridge
Univ. Press), p. 82

\bibitem[]{}  Bolton, C. T. 1994, Ap. \& Sp. Sci., 221, 95 

\bibitem[]{} Crowther, P. A., \& Lennon, D. J. 2006,, A\&A, 446, 279

\bibitem[]{} 
Groote, D. 2003, Magnetic Fields in O, B, and A Stars," ed. L. Balona et al.
ASP Conf. Ser., 205, 243

\bibitem[]{} Groote, D. 2007, priv. commun.


\bibitem[]{} Groote, D., \& Hunger, K. 1976, A\&A, 52, 303 (GH76)

\bibitem[]{} Groote, D., \& Hunger, K. 1977, A\&A, 56, 129 (GH77)

\bibitem[]{} Groote, D., \& Hunger, K. 1982, A\&A, 116, 64

\bibitem[]{} Groote, D., \& Hunger, K. 1997, A\&A, 319, 250

\bibitem[]{} Groote, D., \& Schmitt, J. H. 2004, A\&A, 418, 235

\bibitem[]{} Hesser, J. E., Moreno, H., \& Ugarte, P. 1977, ApJ, 210, L31

\bibitem[]{} Hubeny, I. 1996. \& Heap, S. R. 1996, ApJ, 470 1144

\bibitem[]{} Hubeny, I., Lanz, T., \& Jeffery, S. 1994, Newslett. Anal. Astron. Spetra, 20, 30

\bibitem[]{} Hunger, K., Heber, U., \&  Groote, D.,  1989, A\&A, 224, 57 

\bibitem[]{} Inglis, D. R., \& Teller, E. 1939, ApJ, 90, 439

\bibitem[]{}  Khokhlova, V. L., Vasilchenko, D. V., Stepanov, V. V. \&
Romanyuk, I. I. 2000, Ast. Let., 26, 177 

\bibitem[]{} Krti\v{c}ka, J., Kub\'at, J., \& Groote, D. 2006, A\&A,
460, 145

\bibitem[]{} Krti\v{c}ka, J., Mikul\'{a}\v{s}ek, Z., Zverko, J., \& 
\v{Z}i\v{z}\v{n}ovsk\'{y}, A\&A, 470, 1089

\bibitem[]{} Kurucz R.L., 1993, ATLAS9 Stellar Atmospheres and 2
km\,s$^{-1}$ Grids, Kurucz CD-ROM \#13

\bibitem[]{} Landstreet, J. D., \& Borra, E. F. 1978, ApJ, 224, L5

\bibitem[]{} Lanz, T., \& Hubeny, I. 2007, ApJS, 169, 83

\bibitem[]{} Lesh, J. R. 1968, ApJS, 17, 371

\bibitem[]{} Oksala, M., \& Townsend, R. H. D. 2007, Active OB stars:
Laboratories for Stellar \& Circumstellar Physics, ed. S. Stefl et al., 
ASP Conf. Ser., 361, 476

\bibitem[]{} Osmer, P., \& Peterson, D. M. 1974, ApJ, 187, 117

\bibitem[]{} Preuss, O., Sch\"ussler, M., Holzwarth, V., et al.
2004, A\&A, 417, 987

\bibitem[]{} Reiners, A., Stahl, B., Wolf, B., et al. 2000, A\&A, 363, 585
  (R00)

\bibitem[]{} Sanz-Forcada, J., Franciosini, E., \& Pallavicini, R. 204,
A\& A, 421, 715

\bibitem[]{}  Shore, S. N. 1987, AJ, 94, 731 

\bibitem[]{} Short, C. I., \& Bolton, C. T. 1994, Pulsation, Rotation, \& 
Mass Loss in Early-Type Stars, ed. L. Balona et al., (Noordwijk: Kluwer), 
p. 171

\bibitem[]{} Smith, M. A., \& Groote, D. 2001, A\&A, 372, 208 (SG01)

\bibitem[]{} Smith, M. A., Wade, G. A., et al. 2006, A\&A, 458, 569

\bibitem[]{} Townsend, R. D. T., \& Owocki, S. P. 2005, MNRAS, 357, 215 (TO05)

\bibitem[]{} Townsend, R. D. T., Owocki, S. P., \& Groote, 2005, ApJ, 630, L81
             (TOG05)

\bibitem[]{} Townsend, R. D. T. 2007, priv. commun.

\bibitem[]{} ud-Doula, A., Townsend, R. H. D., \& Owocki, S. P. 2006,
ApJ, 640, L191

\bibitem[]{} Walborn, N. R. 1974, ApJ, 191, L95

\end{thebibliography}
\end{document}